\documentclass[preprint,showpacs,preprintnumbers,amsmath,amssymb,amssymb]{revtex4}
\usepackage{graphicx}
\usepackage{dcolumn}
\usepackage{bm}
\begin{document}

\title{Searching for doubly charged Higgs bosons at the LHC in a
3-3-1 Model}
\author{J.\ E.\ Cieza Montalvo$^1$, Nelson V. Cortez Jr.$^2$,
J. S\'a Borges$^1$ and Mauro D. Tonasse$^3$}
\affiliation{$^1$ Instituto de F\'{\i}sica, Universidade do Estado do Rio
de Janeiro, Rua S\~ao Francisco Xavier 524, 20559-900 Rio de Janeiro,
RJ, Brazil\\
$^2$ Rua Justino Boschetti 40, 02205-050 S\~ao Paulo, SP, Brazil\\
$^3$ {\it Campus} Experimental de Registro, Universidade Estadual Paulista, Rua
Tamekishi Takano 5, 11900-000, Registro, SP, Brazil}
\date{\today}


\begin{abstract}
Using a peculiar version of  the SU(3)$_L\otimes$U(1)$_N$ electroweak model, we investigate the production  of doubly charged Higgs boson at the Large Hadron Collider. Our results include branching ratio calculations for the doubly charged Higgs and for one of the neutral scalar bosons of the model.
\end{abstract}
\maketitle

\section{Introduction \label{sec1}}

In the last few years we have seen a tremendous experimental progress in
the realm of the weak interactions.  However, these advances do not
attain the scalar sector yet. This is the sense in which LHC (Large
Hadron Collider) facilities may shed some light especially on
the Higgs boson. One of  the main ingredients of the Standard Model
is the Higgs mechanism which, in principle, explains how the
particles gain masses through the introduction of an isodoublet
scalar field. The scalar field is the responsible for the
spontaneous breakdown of the gauge symmetry. After electroweak
symmetry breaking, the interaction of this scalar with the gauge
bosons and fermions generate the mass of the particles. In this
process there remains a single neutral scalar, manifesting itself as
the Higgs particle. \par
The Standard Model possibly is a low energy effective theory which
must be generalized by some GUT (Grand Unified Theory). However,
there are several motivations to extend the electroweak theory below
the GUT scale. Supersymmetric Models, for example, provide a
solution to the hierarchy problem through the cancelation of the
quadratic divergences {\it via} fermionic and bosonic loop
contributions \cite{supers}. The Little Higgs Model, recently
proposed, predicts that the Higgs boson is a pseudo-Goldstone boson
of some approximate broken global symmetry \cite{arkani}. Therefore,
this model is also able to solve the naturalness problem of the
Higgs mass.  One of the main motivations for the Left-Right Models
and for the study of their phenomenological consequences is that in
which the Higgs triplet representation furnishes a seesaw
type neutrino mass matrix associated with the presence of a doubly charged Higgs boson \cite{pati}. Therefore, these models suggest a route
to understanding the pattern of the neutrino masses. \par
Doubly charged Higgs bosons appear in several popular extensions of the Standard
Model such as left-right symmetric models \cite{pati} and Higgs
triplet models \cite{georgi}. However, there is another interesting class of electroweak models which also predict particles like that. This class of models is called 3-3-1 Models \cite{PP92,PT93a}. This is the simplest chiral extension of
the Standard Model. It is able to solve the fermion family's
replication problem through of a simple relation between the number of colors and the anomaly cancelation mechanism. It is important to
notice that a solution to this problem is not furnished even in the
context of the GUTs. The 3-3-1 Models have other interesting
features, as for example the upper bound on the Weinberg  mixing
angle, through the relation $\sin^2\theta_W < 1/4$. This feature
does not happen in any kind of others electroweak models except
GUTs, where the value of $\sin^2\theta_W$ is predicted. This result
leads to an upper bound for the energy scale of the model when this
parameter is evolved to high values \cite{DM05}. In a similar
fashion as occurs in Left-Right Model, the seesaw mechanism can be
naturally incorporated in some versions of the 3-3-1 Models
\cite{CT05}. \par 
No Higgs bosons have yet been found. In the meantime,
it is the last brick that is lacking to finish the construction of
the building of the standard electroweak theory. Besides, it is possible that
the Higgs sector brings to light a non standard physics.\par
Since that the 3-3-1 Models are good candidates for physics beyond the
Standard Model, it is interesting to evaluate if the future
accelerators will produce events in sufficient numbers to detect
some of the 3-3-1 Higgs bosons. In particular, there is an increasing
interest in the phenomenology associated with doubly charged Higgs
bosons, a kind that appears in models that admit scalars in triplet
representation of the gauge group \cite{AM05}. Here we are
interested in one of such version of the 3-3-1 Models for which the
scalar fields come only in triplet representation \cite{PT93a,CT05}.
It predicts three new neutrals, four single charged and two doubly
charged Higgs bosons. These scalars can be disclosed in relatively
low energies, which make them interesting for searches in the next
generation of particle accelerators. \par 
Our work is organized as follows. In Sec. \ref{sec2} we summarize the relevant
features of the model, in Sec. \ref{sec3} we present the
cross-section calculations and in Sec. \ref{sec4} we give our
conclusions.

\section{Overview of the Model \label{sec2}}

The underlying electroweak symmetry group is
SU(3)$_L$$\otimes$U(1)$_N$, where $N$ is the quantum number of the U(1) group. Therefore, the left-handed lepton matter content is $\left(\begin{array}{ccc} \nu^\prime_a & \ell^\prime_a &
{\tt L}^\prime_a\end{array}\right)^{\tt T}_L$ transforming as $\left({\bf 3},
0\right)$, where $a = e, \mu, \tau$ is a family index (we are using
primes for the interaction eigenstates). ${\tt L}^\prime_{aL}$ are lepton
fields which can be the charge conjugates ${\ell^\prime_{aR}}^C$
\cite{PP92}, the antineutrinos ${\nu^\prime_{La}}^C$ \cite{MP93} or heavy leptons $P^{\prime+}_{aL}$ $\left(P^{\prime+}_{aL} =
E^{\prime+}_{L}, M^{\prime+}_{L}, T^{\prime+}_{L}\right)$
\cite{PT93a}.  \par 
The model of Ref. \cite{PT93a} has the
simplest scalar sector for 3-3-1 Models. In this version the charge operator is given by 
\begin{equation}
\frac{Q}{e} = \frac{1}{2} \left(\lambda_3 - \sqrt{3}\lambda_8 \right)+ N ,
\label{carga}
\end{equation}\noindent
where $\lambda_3$ and $\lambda_8$ are the diagonal Gell-Mann matrices and $e$ is the elementary electric charge. The right-handed charged leptons are introduced in singlet
representation of SU(3)$_L$ as $\ell^{\prime -}_{aR} \sim \left({\bf
1}, -1\right)$ and $P^{\prime +}_{aR} \sim \left({\bf 1}, 1\right)$.\par
The quark sector is given by
\begin{equation}
Q_{1L} = \left(\begin{array}{c} u^\prime_1 \cr d^\prime_1 \cr J_1
\end{array}\right)_L \sim \left({\bf 3}, \frac{2}{3}\right), \qquad
Q_{\alpha L} = \left(\begin{array}{c} d^\prime_\alpha \cr
u^\prime_\alpha \cr J^\prime_\alpha \end{array}\right)_L \sim
\left({\bf 3}^*, -\frac{1}{3}\right),
\end{equation}
where $\alpha = 2, 3$, $J_1$ and $J_\alpha$ are exotic quarks with electric charge $\frac 5 3$ and $-\frac 4 3$ respectively. It must be notice that the first quark family transforms differently from the two others under the gauge group, which is essential for the anomaly cancelation mechanism \cite{PP92}. \par
The physical fermionic eigenstates rise by the transformations
\begin{subequations}\begin{eqnarray}
&& \ell^{\prime -}_{aL,R} = A^{L,R}_{ab}\ell^-_{bL,R}, \quad
P^{\prime +}_{aL,R} = B^{L,R}_{ab}P^+_{bL,R}, \\ && U^\prime_{L, R}
= {\cal U}^{L, R}U_{L, R}, \quad D^\prime_{L, R} = {\cal D}^{L,
R}D_{L, R}, \quad J^\prime_{L, R} = {\cal J}^{L, R}J_{L, R},
\end{eqnarray}\label{eigl}\end{subequations}
where $U_{L, R} = \left(\begin{array}{ccc} u & c & t
\end{array}\right)_{L, R}$, $D_{L, R} = \left(\begin{array}{ccc} d &
s & b \end{array}\right)_{L, R}$, $J_{L, R} =
\left(\begin{array}{ccc} J_1 & J_2 & J_3 
\end{array}\right)_{L, R}$ and $A^{L, R}$, $B^{L, R}$, ${\cal U}^{L, R}$, ${\cal D}^{L, R}$, ${\cal J}^{L, R}$ are arbitrary mixing matrices. \par 
The minimal scalar
sector contains the three scalar triplets
\begin{equation}
\eta = \left(\begin{array}{c} \eta^0 \\ \eta_1^- \\ \eta_2^+
\end{array}\right) \sim \left({\bf 3}, 0\right), \quad \rho =
\left(\begin{array}{c} \rho^+ \\ \rho^0 \\ \rho^{++}
\end{array}\right) \sim \left({\bf 3}, 1\right), \quad \chi =
\left(\begin{array}{c} \chi^- \\
\chi^{--} \\ \chi^0 \end{array}\right) \sim \left({\bf 3}, -1\right).
\label{eigh}
\end{equation}
The most general, gauge invariant and renormalizable Higgs potential, which conserves the leptobaryon number \cite{PT93b}, is 
\begin{eqnarray}
V\left(\eta, \rho, \chi\right) & = & \mu_1^2\eta^\dagger\eta +
\mu_2^2\rho^\dagger\rho + \mu_3^2\chi^\dagger\chi +
\lambda_1\left(\eta^\dagger\eta\right)^2 +
\lambda_2\left(\rho^\dagger\rho\right)^2 +
\lambda_3\left(\chi^\dagger\chi\right)^2 + \cr && +
\left(\eta^\dagger\eta\right)
\left[\lambda_4\left(\rho^\dagger\rho\right) +
\lambda_5\left(\chi^\dagger\chi\right)\right] + \lambda_6
\left(\rho^\dagger\rho\right)\left(\chi^\dagger\chi\right) +
\lambda_7\left(\rho^\dagger\eta\right)\left(\eta^\dagger\rho\right)
+ \cr && +
\lambda_8\left(\chi^\dagger\eta\right)\left(\eta^\dagger\chi\right)
+
\lambda_9\left(\rho^\dagger\chi\right)\left(\chi^\dagger\rho\right)
+ \frac{1}{2}\left(f\epsilon^{ijk}\eta_i\rho_j\chi_k + {\mbox{H.
c.}}\right) \label{pot}\end{eqnarray} 
The neutral components of the scalars triplets (\ref{eigh}) develop non zero vacuum spectation values $\langle\eta^0\rangle = v_\eta$, $\langle\rho^0\rangle = v_\rho$ and
$\langle\chi^0\rangle = v_\chi$, with $v_\eta^2 + v_\rho^2 = v_W^2 =
(246 \mbox{ GeV})^2$. This mechanism  generate the fermion and gauge boson masses \cite{TO96}. The pattern of symmetry breaking is $\mbox{SU(3)}_L
\otimes\mbox{U(1)}_N\stackrel{\langle\chi\rangle}{\longmapsto}
\mbox{SU(2)}_L\otimes\mbox{U(1)}_Y\stackrel{\langle\eta,
\rho\rangle}{\longmapsto}\mbox{U(1)}_{\rm em}$. Therefore, we can
expect $v_\chi \gg v_\eta, v_\rho$. In the potential (\ref{pot}) $f$
and $\mu_j$ $\left(j = 1, 2, 3\right)$ are  constants with dimension of mass and the $\lambda_i$ $\left(i =
1, \dots, 9\right)$ are adimensional constants with $\lambda_3 < 0$
and $f < 0$ from the positivity of the scalar masses \cite{TO96}. The $\eta$ and
$\rho$ scalar triplets give masses to the ordinary fermions and
gauge bosons, while the $\chi$ scalar triplet gives masses to the
new fermion and gauge bosons. In this work we are using the
eigenstates and masses (see Appendix \ref{apb}) of the Ref. \cite{TO96}. For others analysis of the 3-3-1 Higgs potential see Ref. \cite{DL06}. \par 
Symmetry breaking is initiated when the scalar
neutral fields are shifted as $\varphi = v_\varphi + \xi_\varphi +
i\zeta_\varphi$, with $\varphi$ $=$ $\eta^0$, $\rho^0$, $\chi^0$.
Thus, the physical neutral scalar eigenstates $H^0_1$, $H^0_2$,
$H^0_3$ and $h^0$ are related to the shifted fields as
\begin{subequations}\begin{equation}
\left(\begin{array}{c} \xi_\eta \\ \xi_\rho \end{array}\right)
\approx \frac{1}{v_W}\left(\begin{array}{cc} c_\omega & s_\omega \\
s_\omega & -c_\omega
\end{array}\right)\left(\begin{array}{c} H^0_1 \\
H^0_2 \end{array}\right), \qquad \xi_\chi \approx H^0_3, \qquad
\zeta_\chi \approx ih^0, \label{eign}\end{equation} 
and in the charged scalar sector we have
\begin{eqnarray}
&& \eta^+_1 = s_\omega H^+_1, \qquad \eta^+_2 = s_\varphi H_2^+,
\qquad \rho^+ = c_\omega H_1^+, \\ && \chi^+ = c_\varphi H^+_2,
\qquad \rho^{++} = s_\phi H^{++}, \qquad \chi^{++} = c_\phi H^{++},
\label{eigc}\end{eqnarray}\label{eig}\end{subequations} 
with the condition that $v_\chi \gg v_\eta$, $v_\rho$ in Eqs. (\ref{eign})
and $c_\omega = \cos\omega = v_\eta/\sqrt{v_\eta^2 + v_\rho^2}$,
$s_\omega = \sin\omega$, $c_\phi = \cos\phi = v_\rho/\sqrt{v_\rho^2
+ v_\chi^2}$, $s_\phi = \sin\phi$, $c_\varphi = \cos\varphi =
v_\eta/\sqrt{v_\eta^2 + v_\chi^2}$, $s_\varphi = \sin\varphi$. The $H_1^0$ Higgs boson in Eq. (\ref{eign}) can be the standard model scalar boson, since its mass (see Appendix \ref{apb}) has no dependence on the spectation value $v_\chi$ \cite{TO96}. \par 
The Yukawa interactions for leptons and quarks are, respectively,
\begin{subequations}\begin{eqnarray}
-{\cal L}_\ell & = & G_{ab}\overline{\psi}_{aL}\ell^\prime_{bR}\rho
+ G^\prime_{ab}\overline{\psi}_{aL}P^\prime_{bR}\chi + {\mbox{H.
c.}}, \label{Ll} \\ 
-{\cal L}_Q & = &
\overline{Q}_{1L}\sum_i\left[G^u_{1i}U^\prime_{iR}\eta +
G^d_{1i}D^\prime_{iR}\rho + \sum_\alpha\overline{Q}_{\alpha
L}\left(F^u_{\alpha i}U^\prime_{iR}\rho^* + F^d_{\alpha
i}D^\prime_{iR}\eta^*\right)\right] + \cr && +
G^j\overline{Q}_{1L}J_{1R}\chi +
\sum_{\alpha\beta}G^j_{\alpha\beta}\overline{Q}_{\alpha
L}J^\prime_{\beta R}\chi^* + {\mbox{H. c.}}
\end{eqnarray}\label{Lq}\end{subequations}
In Eqs. (\ref{Lq}), as before mentioned $a,b = e, \mu, \tau$  and $\alpha = 2,3$. \par 
Beyond the standard particles $\gamma$, $Z$ and $W^\pm$ the model predicts, in the gauge sector, one neutral $\left(Z^\prime\right)$, two single charged $\left(V^\pm\right)$ and two double charged $\left(U^{\pm\pm}\right)$ gauge bosons. The interactions between the gauge and Higgs bosons are given by the covariant derivative
\begin{equation}
{\cal D}_\mu\varphi_i = \partial_\mu\varphi_i -
ig\left(\vec{W}_\mu.\frac{\vec{\lambda}}{2}\right)^j_i\varphi_j -
ig^\prime N_\varphi\varphi_iB_\mu,
\label{cov}\end{equation}
where $N_\varphi$ are the U(1) charges for the $\varphi$ Higgs
triplets $\left(\varphi = \eta, \rho, \chi\right)$. $\vec{W}_\mu$
and $B_\mu$ are field tensors of SU(2) and U(1), respectively,
${\vec{\lambda}}$ are Gell-Mann matrices and $g$ and $g^\prime$ are
coupling constants for SU(2) and U(1), respectively. \par 
Introducing the eigenstates (\ref{eigl}) and (\ref{eig}) in the Lagrangeans (\ref{Lq}) we obtain the Yukawa interactions as function of the physical eigenstates, {\it i. e.},
\begin{subequations}\begin{eqnarray}
-{\cal L}_\ell & = &
\frac{1}{2}\left\{\frac{1}{v_\rho}\left[c_\omega\overline{\nu}{\cal
U}^{\nu e}H^+_1 + \left(v_\rho + s_\omega H_1^0 - c_\omega
H_2^0\right)\overline{e^-} + s_\phi \overline{P^+}{\cal
U}^{Pe}H^{++}\right]M^eG_Re^- + \right. \cr && \left. +
\frac{1}{v_\chi}\left[c_\omega\overline{\nu}{\cal V}^{\nu P}H_2^- +
c_\phi\overline{e^-}{\cal V}^{eP}H^{--} + \left(v_\chi + H_3^0 +
ih^0\right)\overline{P^+}\right]M^EG_RP^+\right\} + \cr && +
{\mbox{H. c.}}, \label{llep}\\ 
-{\cal L}_Q & = &
\frac{1}{2}\left\{\overline{U}G_R\left[1 + \left[\frac{s_\omega}{v_\rho}
+ \left(\frac{c_\omega}{v_\eta} +
\frac{s_\omega}{v_\rho}\right){\cal V}^u\right]H_1^0 + \left[-
\frac{c_\omega}{v_\rho} + \left(\frac{s_\omega}{v_\eta} -
\frac{c_\omega}{v_\rho}\right){\cal V}^u\right]H_2^0\right]M^uU +
\right. \cr && \left. + \overline{D}G_R\left[1 +
\left[\frac{c_\omega}{v_\eta} + \left(\frac{s_\omega}{v_\rho} -
\frac{c_\omega}{v_\eta}\right){\cal V}^D\right]H_1^0 +
\left[\frac{s_\omega}{v_\eta} - \left(\frac{c_\omega}{v_\rho} +
\frac{s_\omega}{v_\eta}\right){\cal V}^D\right]H_2^0\right]M^dD +
\right. \cr && + \left. \overline{U}G_R\left[\frac{s_\omega}{v_\eta}V^\dagger_{\rm CKM}H^-_1
+ \left(\frac{c_\omega}{v_\eta} -
\frac{s_\omega}{v_\rho}\right){\cal V}^{ud}H^+_1\right]M^dD +
\right. \cr && + \left.
\overline{D}G_R\left[\frac{c_\omega}{v_\rho}V_{\rm CKM}H^+_1 +
\left(\frac{s_\omega}{v_\eta} - \frac{c_\omega}{v_\rho}\right){\cal
V}^{ud\dagger}H_1^-\right]M^uU\right\} + {\mbox{H. c.}}, \label{lqua}\\ 
-{\cal
L}_J & = & \frac{1}{2}\left[\overline{J}G_R{\cal
J}^{L\dagger}\left({\cal NU}^LM^uU + {\cal RD}^LM^dD\right) +
\right. \cr && \left. + \left(\overline{U}{\cal U}^{L\dagger}{\cal
X}_1 + \overline{D}{\cal D}^{L\dagger}{\cal X}_2 + \overline{J}{\cal
J}^{L\dagger}{\cal X}_0\right){\cal J}^LM^JG_RJ\right] + {\mbox{H.
c.}}, \label{yukj}
\end{eqnarray}\label{yukt}\end{subequations}
where $G_R = 1 + \gamma_5$, $V_L^UV_L^D = V_{\rm CKM}$, the
Cabibbo-Kobayashi-Maskawa mixing matrix, ${\cal U}^{\nu e}$, ${\cal
U}^{Pe}$, ${\cal V}^{\nu e}$, ${\cal V}^{eP}$, ${\cal V}^u =
V_L^U\Delta V_L^{U\dagger}$, ${\cal V}^d = V_L^D\Delta
V_L^{D\dagger}$ and ${\cal V}^{ud} = V_L^U\Delta V_L^{D\dagger}$ are arbitrary mixing
matrices, $M^e = {\mbox{diag}}\left(\begin{array}{ccc} m_e & m_\mu &
m_\tau \end{array}\right)$, $M^P =
{\mbox{diag}}\left(\begin{array}{ccc} m_E & m_M & m_T
\end{array}\right)$, $M^u = {\mbox{diag}}\left(\begin{array}{ccc}
m_u & m_c & m_t \end{array}\right)$, $M^d =
{\mbox{diag}}\left(\begin{array}{ccc} m_d & m_s & m_b
\end{array}\right)$ and $M^J = {\mbox{diag}}\left(\begin{array}{ccc}
m_{J_1} & m_{J_2} & m_{J_3} \end{array}\right)$. In Eq. (\ref{yukj})
we have defined
\begin{subequations}\begin{eqnarray}
{\cal N} = \left(\begin{array}{ccc} s_\omega H_2^+/v_\eta & 0
&             0        \cr
                              0            & s_\phi H^{--}/v_\rho & s_\phi H^{--}/v_\rho \cr
                              0            & s_\phi H^{--}/v_\rho & s_\phi H^{--}/v_\rho \end{array}\right), \quad  {\cal X}_0 \approx \frac{v_\chi + H_3^0 + ih^0}{v_\chi}\left(\begin{array}{ccc} 1 & 0 & 0 \cr
                                                          0 & 1 & 1 \cr
                                                          0 & 1 & 1 \end{array}\right), &&\\
{\cal R} = \left(\begin{array}{ccc}   s_\phi H^{++}/v_\rho   &           0       &        0          \cr
                 0         & s_\omega H_2^-/v_\eta & s_\omega H_2^-/v_\eta \cr
                 0         & s_\omega H_2^-/v_\eta & s_\omega H_2^-/v_\eta \end{array}\right), \quad {\cal X}_1 = \frac{1}{v_\chi}\left(\begin{array}{ccc} c_\omega H_2^- &        0      &      0       \cr
              0          & c_\phi H^{++} & c_\phi H^{++} \cr
              0          & c_\phi H^{++} & c_\phi H^{++} \end{array}\right), &&\\
{\cal X}_2 = \frac{1}{v_\chi}\left(\begin{array}{ccc} c_\phi H^{--} &         0       &        0        \cr
              0        & c_\omega H_2^+ & c_\omega H_2^+ \cr
              0        & c_\omega H_2^+ & c_\omega H_2^+ \end{array}\right).&&
\end{eqnarray}\label{lqj}
\end{subequations} 
We call attention to the fact that non standard field interactions violate leptonic number, as can be seen from the Lagrangians (\ref{Lq}) and (\ref{cov}). However the total leptonic number is conserved \cite{PP92}.

\section{Cross section production \label{sec3}}

The main mechanism for the production of Higgs particles in $pp$
collisions occurs through the mechanism of Drell-Yan and gluon-gluon fusion as shown in Fig. \ref{fig1}. In all calculations  we are
considering that the charged fermionic mixing matrices [see Eqs.
(\ref{eigl})] are diagonals. Using the interaction Lagrangians
(\ref{pot}) and (\ref{yukt}) we first evaluate the differential
cross section for the Drell-Yan process, {\it i. e.}, $pp \to
H^{++}H^{--}$ through the exchange of
$\gamma$, Z, $Z^\prime$, $H_{1}^{0}$ and $H_{2}^{0}$ in the
$s$-channel. Therefore we obtain the differential cross
section for this reaction as
\begin{eqnarray}
\frac{{d\hat\sigma}}{d\Omega} = \frac{1}{64\pi^2\hat s}
\left(\overline{\left|{\cal M}_\gamma\right|^2} +
\overline{\left|{\cal M}_{H_1^0}\right|^2} +
\overline{\left|{\cal M}_{H_2^0}\right|^2} +
\overline{\left|{\cal M}_Z\right|^2} + \overline{\left|{\cal
M}_{Z^\prime}\right|^2}\right. \nonumber
\left. + 2 {\it Re}\overline{{\cal M}_{H_1^0}{{\cal M}_{H_2^0}}^*}
\right),
\end{eqnarray} 
where
\begin{eqnarray}
\frac{d\hat\sigma}{d\cos\theta} & = &
\frac{\beta_{H^{\pm \pm}}}{24}\left\{\frac{\left[\Lambda_1\zeta^{\left(1\right)}\left(\hat
s\right)m_qv_\eta
v_\rho\right]^2 + \left[\Lambda_2 \zeta^{\left(2\right)}\left(\hat
s\right)\right]^2\left(v_\eta^4m_u^2 +
v_\rho^4m_d^2\right)}{\left(2v_Wv_\eta
v_\rho\right)^2}\left[8\left(m_q^2 + m^2_{H^{\pm \pm}}\right) - 4 \hat
u - 4\hat t\right] \right. +  \nonumber \\ 
&&\left. +
\frac{2\pi}{\hat s^3}\left(\frac{\Lambda\alpha
Q_q}{\sin\theta_W}\right)^2\left[\left(\hat s - 2m_q^2\right)\hat s
- 4m^2_{H^{\pm \pm}}\left(\hat s + 2 m_q^2\right) - \left(\hat t - \hat
u\right)^2\right]\right\} + \cr
&& + \sum_{Z,Z^\prime} \frac{\beta_{H^{\pm \pm}}\alpha^{2}\pi{\Lambda^2_{Z\left(Z^\prime\right)}}}{36\sin^2\theta_W\cos^2\theta_W\hat{s}\left(\hat s -
m_{Z\left(Z^\prime\right)}^2 + im_{Z\left(Z^\prime\right)}\Gamma_{Z\left(Z^\prime\right)}\right)^2}\left[8m_{H^{\pm \pm}}^4\left({g_V^{q\left(q^\prime\right)}}^2 + {g_A^{q\left(q^\prime\right)}}^2\right) + \right. \nonumber  \\ 
&& \left. + 8m_{H^{\pm \pm}}^2\left({2m_q}^2 - \hat t - \hat u\right)\left({g_V^{q\left(q^\prime\right)}}^2 + {g_A^{q\left(q^\prime\right)}}^2\right) + 8m_q^4
\left({g_V^{q\left(q^\prime\right)}}^2 + {g_A^{q\left(q^\prime\right)}}^2\right) + \right. \nonumber  \\ 
&& \left. - 8m_q^2\left(\hat t + \hat u\right)\left({g_V^{q\left(q^\prime\right)}}^2  +
{g_A^{q\left(q^\prime\right)}}^2\right) + 8m_q^2\hat s{g_A^{q\left(q^\prime\right)}}^2 + 2\left(\hat t + \hat u\right)^2\left({g_V^{q\left(q^\prime\right)}}^2 + {g_A^{q\left(q^\prime\right)}}^2\right) + \right. \nonumber  \\ 
&&\left.
- 2\hat s^2\left({g_V^{q\left(q^\prime\right)}}^2 + {g_A^{q\left(q^\prime\right)}}^2\right)\right].
\end{eqnarray}
Here, $\sqrt{\hat s}$ is the CM (center of mass) energy of the $q
\bar{q}$ system. For the Standard Model parameters we assume PDG
values, {\it i. e.}, $m_Z = 91.19$ GeV, $\sin^2{\theta_W} = 0.2315$,
and $m_W = 80.33$ GeV \cite{Cea98}. $\Gamma_{Z(Z^\prime)}$ are the total
width of the boson $Z(Z^\prime)$ \cite{cieton1,cieto}. The velocity of the
Higgs in the CM of the process is denoted through
$\beta_{H^{\pm \pm}}$. The $\Lambda_i$ $\left(i = 1, 2\right)$ are the vertex strengths for
$H_1^0H^{--}H^{++}$ and $H_2^0H^{--}H^{++}$, respectively, while
${\Lambda_{\gamma}}_{\mu}$ is one for $\gamma H^{--} H^{++}$ and the
${\Lambda_{Z\left(Z^\prime\right)}}_{\mu}$ is for the bosons
$Z\left(Z^{\prime}\right)H^{--}H^{++}$. The analytical expressions for these vertex strengths are
\begin{subequations}\begin{eqnarray}
\Lambda_1 & = & -2i\left\{2\left[\left(\lambda_6 +
\lambda_9\right)c_\phi^2 + \left(2\lambda_2 +
\lambda_9\right)s_\phi^2\right]s_\omega v_\rho +
c_\omega\left[2\left(\lambda_5c_\phi^2 +
\lambda_4s_\phi^2\right)v_\eta + \right.\right. \cr && \left.\left.
- fc_\phi s_\phi\right]\right\}, \\ \Lambda_2 & = &
ic_\omega\left\{2\left[\left(-\lambda_5 + \lambda_6 +
\lambda_9\right)c_\phi^2 + 2\left(2\lambda_2 - \lambda_4 -
\lambda_9\right)s_\omega\right]v_\eta + fc_\phi s_\phi\right\}, \\
\Lambda_{\gamma\mu} & = & -e\left(c_\phi^2 - s_\phi^2\right)\left(p
- q\right)_\mu, \\ {\Lambda_{Z}}_{\mu} & = &
-e\frac{4\sin^2\theta_W\left(v_\eta^2 - v_\chi^2\right) -
v_\eta^2}{4\left(v_\eta^2 +
v_\chi^2\right)\sin\theta_W\cos\theta_W}\left(p - q\right)_\mu, \\
{\Lambda_{Z^\prime}}_{\mu} & = & -e\frac{\left(10\sin^2\theta_W -
1\right)v_\eta^2 + \left(1 -
7\sin^2\theta_W\right)v_\chi^2}{4\left(v_\eta^2 +
v_\chi^2\right)\sin\theta_W\sqrt{3\cos^2\theta_W\left(1 -
4\sin^2\theta_W\right)}}\left(p - q\right)_\mu.
\end{eqnarray}\end{subequations}
The Higgs parameters $\lambda_i$ $\left(i = 1 \dots 9\right)$ must run from $-3$ to $+3$ in order to allow perturbative calculations. For $H_\alpha^0$ $\left(\alpha = 2, 3\right)$ we take $m_{H_\alpha^0} = \left(0.2 - 3.0\right)$ TeV, while we assume $m_{H_1^0} = 230$ GeV for the standard model scalar one. It must be notice that here there is no contribution from the interference between the  scalar particle $H_{1}^{0}$ and a vectorial one $\left(\gamma, Z
 {\mbox{ or }} Z^\prime\right)$. The kinematic invariant $\hat{t}$ and $\hat{u}$ are,
\begin{subequations}
\begin{align}
\hat t & =  m_{q}^{2} + m_{H_{\pm \pm}}^{2} - \frac{\hat s}{2} \left(1- \cos \theta  \sqrt { 1- \frac{4m_q^2}{\hat s}}\ \sqrt{1- \frac{4 m^2_{H^{\pm \pm}}}{\hat s}}\right),
\\
\hat u & =  m_{q}^{2} + m_{H_{\pm \pm}}^{2} - \frac{\hat s}{2} \left(1+ \cos \theta  \sqrt { 1- \frac{4m_q^2}{\hat s}}\ \sqrt{1- \frac{4 m^2_{H^{\pm \pm}}}{\hat s}}\right).
\end{align}
\label{rs}
\end{subequations}
The total cross section ($\sigma$) for the process $pp \to H^{++} H^{--}$ is related to the cross section ($\hat\sigma$) of the subprocess $qq \to H^{++}H^{--}$ through 
\begin{equation}
\sigma = \sum_{q_i=u,d,s,c}\, \int_{\tau_{min}}^1
\int_{\ln{\sqrt{\tau_{min}}}}^{-\ln{\sqrt{\tau_{min}}}}d\tau dy \
f_{q_{i}} \left(\sqrt{\tau}e^y\right) \
f_{\bar{q}_i}\left(\sqrt{\tau}e^{-y}\right)\hat\sigma\left(\tau,
\hat s\right), 
\end{equation}
where $\tau_{min} = \hat{s}/s$ and $f_{qi}$ and $f_{\bar{q}i}$ are the structure functions of the quark and antiquark in the proton, for which the factorization scale is taken equal to the center of mass energy of the $q\bar{q}$ system and used in our numerical calculations. \par
Another form to produce a pair of doubly charged Higgs is {\it via}
the gluon-gluon fusion through the reaction $gg \to
H^{++}H^{--}$ (see Fig. 1b). As the final state is neutral, the $s$-channel
involves the exchange of the three neutral Higgs $H_1^0$,
$H_2^0$ and $H_3^0$. The exchange of a photon is not allowed by $C$
conservation (Furry's theorem). Therefore, the differential cross
section for gluon-gluon fusion can be expressed by
\begin{eqnarray}
\frac{d\hat\sigma}{d\Omega} & = &
\frac{1}{64\pi^2\hat{s}}\left[\overline{\left|{\cal M}_{H_1^0}
\right|^2} + \overline{\left|{\cal M}_{H_2^0}\right|^2} +
\overline{\left|{\cal M}_{H_3^0}\right|^2} + \right. \cr 
&& \left. + 2\left({\it Re} \overline{{\cal M}_{H_1^0}{\cal M}_{H_2^0}^*} + {\it
Re}\overline{{\cal M}_{H_1^0}{\cal M}_{H_3^0}^*} + {\it Re}\overline{{\cal M}_{H_2^0}{\cal M}_{H_3^0}^*}\right)\right].
\end{eqnarray}
In order to make explicit the different contributions to the
elementary cross section, we will present them separately. The
quark-loop contributions involve the Higgs $H_i$, where $i=1, 2,
3$, which are exchanged in the $s$-channel. For the $H_1^0$ we have
\begin{eqnarray}
\hat\sigma_{H_1^0} & = &
\frac{\beta_{H^{++}}}{8192\pi^3}\left[\alpha_s\Lambda_1
\zeta^{\left(1\right)}\right]^2\left\{\frac{1}{\hat
s}\left|\sum_{q}\Lambda_{q_1} m_q^2I_d\right|^2 + \hat s\left|\sum_{q
}\Lambda_{q_1} \lambda_q\left[2 + \left(4\lambda_q - 1\right)I_q\right]\right|^2 +
\right. \cr 
&& \left.-2\pi\sum_{q} m_q^2 \lambda_q\ln \left(\frac{r_{+q}}{r_{-q}} \right)\left[\ln^2\left(\frac{r_{+q}}{r_{-q}} \right) - \pi^2 \right]  \right\}.
\label{siH1}\end{eqnarray}
We fix the scale parameter $\Lambda = 0.2$ GeV and the appropriate scale where the strong coupling constant $\alpha_s$ is evaluated as being equal to the center of mass energy of the subprocess, both, used in our numerical calculations. The sums in Eq. (\ref{siH1}) run over all generations where $\lambda_q = m_q^2/\hat{s}$ and  $r_{\pm q} = 1 \pm \sqrt{1 - 4 \lambda_q}$. The $\Lambda_{q_1}$ is the $q\overline{q}H_1^0$ vertex strength,
\begin{equation}
\Lambda_{q_1} = -i\frac{m_q}{2v_W}G_R.
\end{equation}
We also define the propagator for Higgs bosons,
\begin{equation}
\zeta^{\left(i\right)} (\hat{s}) = \frac{1}{\hat{s}- m_{H_{i}}^{2}+
i m_{H_{i}}\Gamma_{H_{i}}},
\label{prop}\end{equation}
where $\Gamma_{H_i^0}$ are the total width of the $H_1^0$ and $H_2^0$ boson, with $i = 1$ for $q = u, c, t, d, s, b$ and $i = 2$ for $u(c, t)$ or $d(s, b)$, separately \cite{cieto}. The Eq. (\ref{prop}) defines also the propagator for $H_3^0$ with $i = 3$. However, $H_3^0$ does not contribute to the Drell-Yan process.
\par 
For the Higgs $H_2^0$ we have
\begin{eqnarray} 
\hat{\sigma}_{H_2^0} & = &
\frac{\beta_{H^{++}}}{8192\pi^3}\left[\alpha_s\Lambda_2
\left|\zeta^{\left(2\right)}\left(\hat
s\right)\right|\right]^2\left\{\frac{1}{\hat
s} \sum_{q} \Lambda_{q_2} \left(m_u^2\frac{v_\eta}{v_\rho} +
m_d^2\frac{v_\rho}{v_\eta}\right)^2\left|I_q\right|^2 +
\right.  \nonumber  \\ 
&& \left. + \hat s\left(\frac{v_\eta}{v_\rho}
+ \frac{v_\rho}{v_\eta}\right)^2\left|\sum_{q} \Lambda_{q_2} \lambda_q\left[2 +
\left(4\lambda_q - 1\right)I_q\right]\right|^2 + \right. \cr
&& \left. + 4\pi \sum_{q} \Lambda_{q_2}\ln\left(\frac{r_{+q}}{r_{-q}}\right) \left(-m_u^2\frac{v_\eta^2}{v_\rho^2} + m_d^2\frac{v_\rho^2}{v_\eta^2}\right) \lambda_q\right\},
\end{eqnarray}
where $\Lambda_{q_2}$
is the strength of the $q\overline{q}H_2^0$ vertex, {\it i. e.}:
\begin{equation}
\Lambda_{q_2} = -\frac{i}{2v_W}G_R\left(m_u\frac{v_\eta}{v_\rho} +
m_d\frac{v_\rho}{v_\eta}\right).
\end{equation}
The contribution of $H_3^0$ to the cross-section is
\begin{equation}
\hat{\sigma}_{H_3^0} =
\frac{\beta^{H^{++}}}{8192\pi^3}\left[\alpha_2 \Lambda_3
\left|\zeta^{\left(3\right)}\right|\right]^2
\left|\sum_{J = J_1, J_2, J_3} \Lambda_{J3} \lambda_J\left[2 +
\left(4\lambda_J -1\right)I_J\right]\right|^2,
\label{sh3}\end{equation} 
where $\Lambda_{J3}$ describe the quark vertex with the Higgs $H_3^0$ and $\Lambda_3$ for the $H_3^0$ with $H^{\pm\pm}$, {\it i. e.},
\begin{subequations}\begin{eqnarray}
\Lambda_{J3} & = & -i\frac{m_{J}}{2v_\chi}G_R, \\ \Lambda_3 & = &
-2iv_\chi\left[\left(\lambda_6 - \lambda_9 \right)s_\varphi^2 +
\left(2\lambda_3 + \lambda_9\right)c_\phi^2\right].
\end{eqnarray}\end{subequations}
We note that $H_3^0$ decays only into exotic
leptons and quarks because it becomes  massive at the  first symmetry breaking.
Therefore, the $H_3^0$ total width is obtained from
\begin{eqnarray}
\Gamma\left(H_3^0 \to {\rm all}\right) & = & \Gamma^0_{J_1\bar{J_1}} + \Gamma^0_{J_2\bar{J_2}} +
\Gamma^0_{J_3\bar{J_3}} + 3\Gamma^0_{P^{\pm}P^{\mp}} + \Gamma^0_{H^0_1H^0_2} + \Gamma^0_{H^{\pm}_1H^{\mp}_1} + \cr
&& + \Gamma^0_{H^{\pm}_2H^{\mp}_2} + \Gamma^0_{H_1^{\pm}H_2^{\mp}} + \Gamma^0_{U^{\mp\mp}U^{\pm\pm}},
\label{w1}\end{eqnarray}
where $\Gamma^0_{XY} = \Gamma\left(H_3^0 \to XY\right)$. The partial widths are show in the Appendix \ref{apa}. \par
The total width of the decay of the Higgs $H^{\pm\pm}$ in quarks, leptons, standard charged gauge boson and charged Higgs $\left(W^\pm H_2^\pm\right)$,
single charged Higgs $\left(H_1^\pm H_2^\pm\right)$, doubly charged gauge bosons and a photon
$\left(U^{\pm\pm}\gamma\right)$, doubly charged bosons and Higgs
$\left(U^{\pm\pm}h^0, U^{\pm\pm}H_1^0, U^{\pm\pm}H_2^0,
U^{\pm\pm}H_3^0\right)$, doubly charged bosons and $Z$ or $Z^\prime$
$\left(U^{\pm\pm}Z, U^{\pm\pm}Z^\prime\right)$, and charged extra gauge boson and Higgs
$\left(V^\pm H_1^\pm\right)$ is given by
\begin{eqnarray}
\Gamma\left(H^{\pm\pm} \to {\rm all}\right)& = & \Gamma^{\pm\pm}_{\bar{J_1}q_{d, s, b}} + \Gamma^{\pm\pm}_{\bar{q}_{u, c, t}J_{2, 3}} + \Gamma^{\pm\pm}_{\bar{J}_{2, 3}q_{u, c, t}} +
\Gamma^{\pm\pm}_{e^\pm P^{\pm\pm}} + \Gamma^{\pm\pm}_{W^\pm
H_2^\pm} + \Gamma^{\pm\pm}_{H_1^\pm H_2^\pm} + \Gamma^{\pm\pm}_{U^{\pm\pm}\gamma} + \cr
&& + \Gamma^{\pm\pm}_{U^{\pm\pm}h^0} + \Gamma^{\pm\pm}_{U^{\pm\pm}H_1^0} + \Gamma^{\pm\pm}_{U^{\pm\pm}H_2^0} + \Gamma^{\pm\pm}_{U^{\pm\pm}H_3^0} + \Gamma^{\pm\pm}_{U^{\pm\pm}Z} + \Gamma^{\pm\pm}_{U^{\pm\pm}Z^\prime} + \cr
&& + \Gamma^{\pm\pm}_{V^\pm H_1^\pm} 
\label{w2}\end{eqnarray}
with $\Gamma^{\pm\pm}_{XY} = \Gamma\left(H^{\pm\pm} \to XY\right)$ (see Appendix \ref{apa} for the partial widths). \par 
The contribution for the interference of the $H_1^0$ and $H_2^0$ is given by
\begin{eqnarray}
\frac{d\hat\sigma_{H^0_1-H_2^0}}{d\Omega} & = & \sum_{q_u,q_d}
\frac{\alpha_s^4\Lambda_1\Lambda_2\Lambda_{k_1}\Lambda_{k_2}
\zeta^{\left(2\right)}\left(\hat
s\right)\delta_{ab}}{8\pi}\left\{\frac{\varepsilon_{k_1}^{\mu
\left(a\right)}\varepsilon_{k_2}^{\nu\left(b\right)}k_1^\alpha
k_2^\beta}{\hat s^2}m_q^2 I_q \right. \\
&& \left. + i\frac{\varepsilon_{k_1}^a\varepsilon_{k_2}^b\delta_{ab}}{2}\lambda\left[2
+ \left(4\lambda - 1\right)I_q\right]\right\} \times \cr && \times
\left\{\frac{\varepsilon_{k_1}^{\mu\left(a\right)}\varepsilon_{k_2}^{\nu\left(b\right)}k_1^\alpha
k_2^\beta\varepsilon_{\alpha\mu\nu\beta}}{\hat
s^2}\left(\frac{v_\eta}{v_\rho}m_u^2I_{q_u} -
\frac{v_\rho}{v_\eta}m_d^2I_{q_d}\right) + \right. \cr && \left. +
i\frac{\varepsilon_{k_1}^a\varepsilon_{k_2}^b}{2}\left[\frac{v_\eta}{v_\rho}\lambda_U\left[2
+ \left(4\lambda_U - 1\right)I_{q_u}\right] -
\frac{v_\rho}{v_\eta}\lambda_D\left[2 + \left(4\lambda_D -
1\right)I_{q_d}\right]\right]\right\},
\end{eqnarray}
where $\varepsilon^{\mu,\nu}$ are the polarizations, $k_{1,2}$ are
the gluon momentum vectors, $m_{q_u}$ are the masses of the $u$,
$c$, $t$ quarks (5 MeV,  1.5 GeV and 175 GeV respectively),
$m_{q_d}$ are the masses of the $d$, $s$, $b$ quarks (9 MeV, 150
MeV and 5 GeV respectively) \cite{Cea98}, $\lambda$ is referred to all quarks and $\lambda_U$ for the quark $u, c, t$ and $\lambda_D$ for $d, s, b$ respectively and $\varepsilon_{\alpha\mu\nu\beta}$
is the antisymmetric tensor.   \par 
The contribution for the interference of $H_1^0$ and $H_3^0$ gives
\begin{eqnarray}
\frac{d\hat\sigma_{H_1^0-H_3^0}}{d\Omega} & = & \sum_{q} i\frac{\alpha_s^4\Lambda_1\Lambda_3\Lambda_{q_1}\Lambda_{q_3}\zeta^{\left(1\right)}\left(\hat s\right)\zeta^{\left(3\right)}\delta_{a^\prime b^\prime}\varepsilon_{q_1}^{a^\prime}\varepsilon_{q_2}^{b^\prime}\delta_{ab}}{256\pi^2}\lambda\left[2 + \left(4\lambda - 1\right)I_q\right]\times \cr
&& \times \left\{i\frac{\varepsilon^a_{q_1}\varepsilon^b_{q_2}}{2}\lambda\left[2 + \left(4\lambda - 1\right)I_q\right] - \frac{\varepsilon_{q_1}^{\mu\left(a\right)}\varepsilon_{q_2}^{\left(b\right)}\varepsilon_{\alpha\mu\nu\beta}}{\hat s^2}\right\},
\end{eqnarray}
and finally we have the following for the $H_2^0$ and $H_3^0$ 
\begin{eqnarray}
\frac{d\hat\sigma_{H_2^0-H^0_3}}{d\Omega} & = & \sum_{q}i\frac{\alpha_s^4\Lambda_2\Lambda_3\Lambda_{q_2}\Lambda_{q_3}\zeta^{\left(2\right)}\left(\hat s\right)\zeta^{\left(3\right)}\left(\hat s\right)\delta_{a^\prime b^\prime}\varepsilon_{q_1}^{a^\prime}\varepsilon_{q_2}^{b^\prime}\delta_{ab}}{256\pi^2}\lambda\left[2 + 4\left(4\lambda - 1\right)I_q\right] \times \cr
&& \times \left\{\varepsilon_{q_1}^{\mu\left(a\right)}\varepsilon_{q_2}^{\nu\left(b\right)}\varepsilon_{\alpha\mu\nu\beta}\left(\frac{v_\eta}{v_\rho}m_u^2I_{q_u} - \frac{v_\rho}{v_\eta}m_d^2I_{q_d}\right) + i\frac{\varepsilon_{q_1}^a\varepsilon_{q_2}^b}{2}\left[\frac{v_\eta}{v_\rho}\lambda_u\left[2 + \left(4\lambda_u - 1\right)I_{q_u}\right] + \right.\right. \cr
&& \left.\left. - \frac{v_\rho}{v_\eta}\lambda_d\left[2 + \left(4\lambda_d - 1\right)I_{q_u}\right]\right]\right\}.
\end{eqnarray}
The loop integrals involved in the evaluation of the elementary cross
 section can be expressed in terms of the function $I_q\left(
 \lambda_q\right) \equiv I_q$ which is defined through
\begin{equation}
I_q = \int_0^1\frac{dx}{x}\ln\left[1 - \frac{\left(1 - x\right)x}{
\lambda_q}\right] = \frac{1}{2}\left\{\begin{array}{rr} -
4\arcsin^2\left[1/\left(2\sqrt{\lambda_q}\right)\right], & {\mbox{
}} \lambda_q > 1/4, \cr \left[\ln\left(r_{+q}/r_{-q}\right) +
2i\pi\right]\ln\left(r_{+q}/r_{-q}\right) - \pi^2, & {\mbox{ }} \lambda_q
< 1/4.\end{array}\right.
\end{equation}
Here, $q$ stands for the quarks running in the loop (Fig. \ref{fig1}b). \par 
The total cross section ($\sigma$) for the process $pp \to gg \to
H^{--} H^{++}$ is related to the total cross section ($\hat{\sigma}$)
through the subprocess $gg \to H^{--} H^{++}$, {\it i. e.} 
\begin{equation}
\sigma = \int_{\tau_{min}}^1 \int_{\ln{\sqrt{\tau_{min}}}}^{-\ln{
\sqrt{\tau_{min}}}}d\tau dy G\left(\sqrt{\tau}e^y,
Q^2\right)G\left(\sqrt{\tau}e^{-y},
Q^2\right)\hat{\sigma}\left(\tau,\hat{s}\right),
\end{equation}
where $G(x, Q^{2})$ is the gluon structure function, for which the factorization scale is taken equal to the center of mass energy of the subprocess and used in our numerical calculations.
\section{Results and Conclusions \label{sec4}}

In this work we have calculated the pair production of doubly charged Higgs by computing the contributions due the Drell-Yan and quark loop processes. In Sec. \ref{sec3} we have given the analytical expressions that allow the numerical evaluations of these contributions and it was showed that the dominant
contribution come from the well known Drell-Yan process. We have presented
the cross section for the process $pp \to H^{--}H^{++}$ involving
the Drell-Yan mechanism and the gluon-gluon fusion, to produce such
Higgs bosons at the LHC (14 TeV). \par
Taking  into account that  the masses of the gauge bosons, Higgs and some other parameters must satisfy the limits imposed by the model (see Sec. \ref{sec2}), besides the approximations in the calculations of masses (Appendix \ref{apb}) and eigenstates (Sec. \ref{sec2}) given in Ref. \cite{TO96}, we have considered two possibilities: $f \simeq 0$ and $f = -99.67$ GeV (see Table \ref{tab1}). For both possibilities we have assumed the values $v_\eta$ = 195 GeV, $v_\chi$ = 1300 GeV, $\lambda_1 = -1.2$, $-\lambda_2 = -\lambda_3 = \lambda_6 = -\lambda_8 = 1$, $\lambda_4 = 2.98$, $\lambda_5 = -1.57$, $\lambda_7 = -3$, but for  $\lambda_9$ we have used  $\lambda_9 = -1$ when  $f \approx 0$ and $\lambda_9 = -1.9$ when $f = -99.67$ GeV.

\begin{table}[h]
\caption{\label{tab1} Approximated values of the masses (see Appendix \ref{apb}) used in this work. All the values in this table are given in GeV.}
\begin{ruledtabular}
\begin{tabular}{c|ccccccccccccccc}
$f$ & $m_E$ & $m_M$ & $m_T$ & $m_{H_1^0}$ & $m_{H_2^0}$ & $m_{H_3^0}$ & $m_{h^0}$ & $m_{H^\pm_1}$ & $m_{H^\pm_2}$ & $m_V$ & $m_U$ & $m_{Z^\prime}$ & $m_{J_1}$ & $m_{J_2}$ & $m_{J_3}$ \\
\hline
$\approx 0$ &\raisebox{-1.5ex}{ 194} &\raisebox{-1.5ex}{ 1138} &\raisebox{-1.5ex}{ 2600 }&\raisebox{-1.5ex}{ 874 } & \raisebox{-1.5ex}{ 1322 } & \raisebox{-1.5ex}{ 2600 } & 0  & 426 & 1315 & \raisebox{-1.5ex}{603} & \raisebox{-1.5ex}{ 601 } & \raisebox{-1.5ex}{ 2220 } & \raisebox{-1.5ex}{ 1300 } & \raisebox{-1.5ex}{ 1833 } & \raisebox{-1.5ex}{ 1833}\\
    -99.67  &  &  &  &  &  &  & 520 & 218 & 1295 &  &  &  &  &  & \\
\end{tabular}
\end{ruledtabular}
\end{table}
The masses of the exotic bosons in Table \ref{tab1} are in accordance with the estimated values  of CDF and D0 experiments, which probe their masses in the range from 500 GeV  to  800 GeV \cite{tait}, while the reach of the LHC will be in the range $1 {\mbox{ TeV}} < m_{Z^\prime} \leq 5$ TeV \cite{freitas}. \par
In Fig. \ref{fig2} we show the cross section for the process $pp \to H^{++}H^{--}$ for $f \simeq 0$ GeV. Considering that the expected
integrated luminosity for the LHC will be of order of $3\times 10^5$
pb$^{-1}$/yr then the statistics give a total of $\simeq 2\times 10^{4}$
events per year for Drell-Yan and $\simeq 0.2$ events per year for
gluon-gluon fusion, for $m_{H^\pm} =
1309$ GeV. \par
\begin{table}[h]
\caption{\label{tab2} Branching ratios for the $H_3^0$ decay with $m_{H_3^0}=2600$
GeV. The notation used in this table is $BR^0_{XY} = BR\left(H_3^0 \to XY\right)$.}
\begin{ruledtabular}
\begin{tabular}{c|cccccc}
$f$ (GeV) & $10^{-5}\times BR^0_{H_1^0H_2^0}$ & $10^{-8}\times BR^0_{H_1^+H_1^-}$ & $BR^0_{H_2^+H_2^-}$ & $BR^0_{H_1^+H_2^-}$ & $BR^0_{E^+E^-}$ & $BR^0_{M^+M^-}$ \\
\hline
$\approx 0$ & 3.35  &  2  &   No & \raisebox{-1.5ex}{0.9999} & \raisebox{-1.5ex}{2 $\times$ 10$^{-7}$} &\raisebox{-1.5ex}{2 $\times$ 10$^{-6}$} \\
    -99.67     & 3.14  &  4  & 4 $\times$ 10$^{-7}$ & & & \\
\end{tabular}
\end{ruledtabular}
\end{table}
In Fig. \ref{fig3}, we show the results for the same process when  $f = -99.67$ GeV. 
This value was calculated considering  the exotic boson masses in the range from $500$ GeV to $800$ GeV  and $v_{\eta}$ having a minimum value of $194.2$ GeV, which assure the values of the $\lambda_{i}$ between $-3$ to $+3$, so we obtain the mass of the doubly charged Higgs, {\it i. e.}, $m_{H^{++}}$=$ 1780$ GeV. Considering the same integrated luminosity as above  this gives a total of $585$ events per
year for Drell-Yan and $0.13$ events per year for gluon-gluon
fusion. We present in Table \ref{tab2} the branching ratios for $H_3^0
\to {\rm all}$ with $f \simeq 0$  and $f = -99.67$ GeV and we can
observe that, due to the coupling constant, the largest width corresponds to $H_3^0 \to H_1^+H_2^-$ decay. In Table \ref{tab3} we present the branching ratios for $H^{\pm\pm} \to$ all.
\begin{table}[h]
\caption{\label{tab3} Branching ratio for the $H^{\pm\pm}$ decay with $m_{H^{\pm\pm}} = 1308$ GeV. Here $BR^{\pm\pm}_{XY} = 10^{-3}\times BR\left(H^{\pm\pm} \to XY\right)$.}
\begin{ruledtabular}
\begin{tabular}{c|ccccccc}
$f$ (GeV) & $BR^{\pm\pm}_{\bar J_1q}$ & $BR^{--}_{\ell^- E^-}$ & $BR^{--}_{\ell^- M^-}$ & $BR^{++}_{\ell^+ E^+}$ & $BR^{++}_{\ell^+ M^+}$ & $BR^{\pm\pm}_{U^{\pm\pm}\gamma}$  \\
\hline
$\approx 0$ & 0.001 & 0.08 & 0.005 & 3 & 6 & 29 \\
    -99.67  & 2 & 0.001 & 0.004 & 0.4 & 4 & 2 \\
\hline\hline
$f$ (GeV) & $BR^{\pm\pm}_{W^\pm H_2^\pm}$ & $BR^{\pm\pm}_{V^\pm H_1^\pm}$ & $BR^{\pm\pm}_{H_1^\pm H_2^\pm}$ & $BR^{\pm\pm}_{U^{\pm\pm}H_1^0}$ & $BR^{\pm\pm}_{U^{\pm\pm}Z}$ & $BR^{\pm\pm}_{U^{\pm\pm}h^0}$\\
\hline
$\approx 0$  & No & 19 & No & No & 444 & 2\\
    -99.67 & 0.09 & 13 & 329 & 6 & 146 & 0.5\\
\end{tabular}
\end{ruledtabular}
\end{table}
From Table \ref{tab3} it can also be noticed that, as the branching ratio(BR) for $H^{\pm \pm} \to  H_{1}^{\pm} H_{2}^{\pm}$ and $H^{\pm \pm} \to U^{\pm \pm} Z,\gamma$ are large, these channels could lead to some interesting signal.\par
We emphasize that the window for varying the free parameters is small because of the constraints imposed on the model. In summary, the analysis of the values in Tables \ref{tab1}, \ref{tab2} and \ref{tab3} show that, although a large number of doubly charged Higgs can be produced by the Drell Yan mechanism, the decays of these particles into ordinary quarks and leptons do not lead to a good signature for its detection even for energies which the LHC can reach.

\appendix

\section{Partial Widths \label{apa}} 

In this Appendix we present the partial widths for Higgs decays from the Eqs. (\ref{w1}) and (\ref{w2}). We define
\begin{equation}
R\left(a, b; x\right) = \frac{1}{16 \pi x^{3/2}} \sqrt{\left[ x - (m_a + m_b)^2\right]\, \left[ x - (m_a - m_b)^2 \right]}
\end{equation}
and we obtain the partial widths for $H_3^0$, with $\sqrt {s} = m_{H^0_3}$ as 
\begin{subequations}\begin{eqnarray}
\Gamma^0_{\overline{J}_iJ_i} & = & 3R\left(J_i, J_i; s\right)\left(\frac{m_{J_i}}{v_\chi}\right)^2\left(m^2_{H_3^0} - 2m_{J_i}^2\right), \\ 
\Gamma^0_{P^+P^-} & = &
R\left(P^+, P^-; s\right)\left(\frac{m_P}{v_\chi}\right)^2\left(m^2_{H_3^0} - 2m_P\right), \\ \Gamma^0_{H_1^0H_2^0} & = & R\left(H_1^0, H_2^0; s\right)\left[4\left(\lambda_5 -
\lambda_6\right)c_\omega v_\rho v_\chi + f\left(c_\omega v_\eta - s_\omega v_\rho\right)\right]^2, \\ 
\Gamma^0_{H^-_1H^+_1} & = & R\left(H_1^+, H_1^+; s\right)\left[2v_\chi\left(\lambda_5s_\omega v_\rho + \lambda_6c_\omega v_\eta\right) + fc_\omega v_\rho\right]^2, \\
\Gamma^0_{H_2^-H_2^+} & = & R\left(H_2^+, H_2^+; s\right)\left[-2v_\chi\left(\lambda_5c_\omega v_\eta + \lambda_6s_\omega v_\rho\right) + fs_\omega v_\eta\right]^2, \\
\Gamma^0_{H_1^-H_2^+} & = & R\left(H_1^+, H_2^+; s\right)\left[\left(\lambda_6 - \lambda_5\right)c_\omega v_\rho v_\chi + f\left(s_\omega v_\rho - c_\omega v_\eta\right)\right], \\ 
\Gamma^0_{U^{--}U^{++}} & = & \frac{g^4v_\chi}{2}R\left(U^{\pm\pm}, U^{\pm\pm};
s\right)\left\{3 - \left(\frac{m_{H_3^0}}{m_{U^{++}}}\right)^2\left[1 - \left(\frac{m_{H_3^0}}{2m_{U^{++}}}\right)^2\right]\right\}, \\
\Gamma^0_{H^{--}H^{++}} & = & 4v_\chi R\left(H^{\pm\pm}, H^{\pm\pm}; s\right)\left[\left(\lambda_6 + \lambda_9\right)s_\varphi^2 + \left(2\lambda_3 + \lambda_9\right)c_\varphi^2\right], \label{xx}
\end{eqnarray}\end{subequations}
where $\Gamma^0_{XY} \equiv \Gamma\left(H^0_3 \to XY\right)$. Finally we present the partial widths for the $H^{--}$ decays, with $\sqrt{s} = m_{H^{++}}$,
\begin{subequations}\begin{eqnarray}
\Gamma^{++}_{\overline{J}_1q_{d, s, b}} & = & 3R\left(J_1, q; s\right)\left(\frac{m_qs_\varphi}{v_\eta}\right)^2\left(m^2_{H^{++}} - m^2_{J_1} - m^2_{d, s, b}\right), \\
\Gamma^{++}_{\overline{q}_{u, c, t}J_{2, 3}} & = & 3R\left(J_{2,3}, q; s\right)\left(\frac{m_{J_{2, 3}}c_\varphi}{v_\chi}\right)^2\left(m^2_{H^{++}} - m^2_{J_{2, 3}} - m^2_{u, c, t}\right), \\ 
\Gamma^{++}_{\overline{J}_{2,3}q_{u, c, t}} & = & 3R\left(J_{2,3}, q; s\right)\left(\frac{m^2_{q_{u, c, t}}s_\varphi}{v_\eta}\right)^2\left(m^2_{H^{++}} - m^2_{J_{2, 3}} -
m^2_{u, c, t}\right), \\ 
\Gamma^{++}_{e^-P^-} & = & \frac{R\left(e, P; s\right)}{4}\left(\frac{m_es_\varphi}{v_\eta}\right)^2\left(m^2_{H^{++}} - m_e^2 - m_P^2\right), \\
\Gamma^{++}_{e^+P^+} & = & \frac{R\left(e, P; s\right)}{4}\left(\frac{m_Pc_\varphi}{v_\chi}\right)^2\left(m^2_{H^{++}} - m^2_e - m^2_P\right), \\
\Gamma^{++}_{W^-H_2^-} & = & \frac{R\left(W, H_2^-; s\right)}{32}\left(\frac{ec_\varphi
c_\phi}{\sin\theta_Wm_W}\right)^2\times \cr && \times\left\{\left(m^2_{H^{++}} - m_W^2\right)^2 + m^2_{H_2^+}\left[m^2_{H_2^+} - \left(m^2_{H^{++}} + m^2_W\right)\right]\right\}, \\ 
\Gamma^{++}_{H_1^-H_2^-} & = & R\left(H_2^-, H_2^-; s\right)\left\{\left[\left(\lambda_7 + \lambda_9\right)s_\omega^2 + \left(\lambda_7 + \lambda_8\right)c_\omega^2\right]s_\phi s_\varphi + \right. \cr 
&& \left. +\frac{\left(\lambda_8 - \lambda_9\right)c_\omega c_\phi v_\rho + fs_\omega s_\phi}{\sqrt{v_\eta^2 + v_\chi^2}}\right\}, \\
\Gamma^{++}_{U^{--}\gamma} & = & \frac{3}{4\pi m_{H^{++}}}\left(\frac{e^2c_\varphi
v_\chi}{\sin^2{\theta_W}}\right)^2\left[1 - \left(\frac{m_U{++}}{m_{H{++}}}\right)^2\right], \\
\Gamma^{++}_{U^{--}H_1^0} & = & \frac{R\left(U^{--}, H_1^0; s\right)}{32}\left(\frac{ec_\varphi v_\chi}{m_{U^{++}}v_W\sin\theta_W}\right)^2 \times \cr 
&& \times\left\{\left(m^2_{H^{++}} - m^2_{U^{++}}\right)^2 + m^2_{H_1^0}\left[m^2_{H_1^0} - 2\left(m^2_{H^{++}} + m^2_{U^{--}}\right)\right]\right\}, \\
\Gamma^{++}_{U^{--}H_2^0} & = & \frac{R\left(U^{--}, H_2^0; s\right)}{32}\left(\frac{ec_\phi v_\rho}{m_{U^{++}}v_W\sin\theta_W}\right)^2 \times \cr && \times\left\{\left(m^2_{H^{++}} - m^2_{U^{--}}\right)^2 + m^2_{H_2^0}\left[m^2_{H_1^0} - 2\left(m^2_{H^{++}} + m^2_{U^{++}}\right)\right]\right\}, \\
\Gamma^{++}_{U^{--}H_3^0} & = & \frac{R\left(U^{--}, H_3^0; s\right)}{32}\left(\frac{ec_\varphi}{m_{U^{++}}\sin^2\theta_W}\right)^2 \times \cr 
&& \times \left\{\left(m^2_{H^{++}} - m^2_{U^{++}}\right)^2 + m^2_{H_3^0}\left[m^2_{H^0_3} -
2\left(m^2_{H^{++}} + m^2_{U^{++}}\right)\right]\right\}, \\
\Gamma^{++}_{U^{--}Z} & = & R\left(U^{--}, Z; s\right)\left(\frac{e^2c_\varphi v_\chi}{\cos\theta}\right)^2 \times \cr 
&& \times \left[5 + \frac{\left(m^2_{H^{++}} - m_Z^2\right)^2 + m^2_{U^{++}}\left(m^2_{U^{++}} -
2m^2_{H^{++}}\right)}{2m^2_{U^{++}}m^2_Z}\right], \\
\Gamma^{++}_{U^{--}Z^\prime} & = & \frac{R\left(U^{--}, Z^\prime; s\right)}{12\left(\sin^2\theta_W - 1\right)\left(4\sin^2\theta_W - 1\right)}\left[\frac{ev_\eta v_\chi\left(10\sin^2\theta_W - 1\right)}{\sin^2\theta_W}\right]^2\times \cr && \times \left[5 + \frac{\left(m^2_{H^{++}} - m_{Z^\prime}^2\right)^2 + m^2_{U^{++}}\left(m^2_{U^{++}} - 2m^2_{H^{++}}\right)}{2m^2_{U^{++}}m^2_{Z^\prime}}\right], \\
\Gamma^{++}_{V^-H_1^-} & = & R\left(V^-, H_1^-; s\right)\left(\frac{ev_\rho s_\varphi}{v_W\sin\theta_W}\right)^2 \times \cr && \times \left\{\left(m^2_{H^{++}} - m^2_V\right)^2 + m^2_{H_1^0}\left[m^2_{H_1^+} - 2\left(m^2_{H^{++}} + m_V^2\right)\right]\right\}.
\end{eqnarray}\end{subequations}
Here, $\Gamma^{++}_{XY} \equiv \Gamma\left(H^{++} \to XY\right)$.

\section{3-3-1 Particle Masses \label{apb}}

In this Appendix we present the expressions of gauge, Higgs boson and fermion masses predicted in 3-3-1 energy scale in terms of the VEVs and the parameters of the potential. 
\begin{subequations}\begin{eqnarray}
&& m^2_{H^0_1} \approx 4\frac{\lambda_2v_\rho^4 - 2\lambda_1v_\eta^4}{v_\eta^2 - v_\rho^2}, \qquad m^2_{H^0_2} \approx \frac{v_W^2}{2 v_\eta v_\rho^2}v_\chi^2, \qquad m^2_{H^0_3} \approx
-4\lambda_3v_\chi^2, \\
&& m^2_h = -\frac{fv_\chi}{v_\eta v_\rho}\left[v_W^2 + \left(\frac{v_\eta v_\rho}{v_\chi}\right)^2\right], \qquad m_{H_{1}^\pm}^2 = \frac{v_W^2}{2 v_\eta v_\rho}\left(fv_\chi - 2\lambda_7v_\eta v_\rho\right), \\
&& m_{H_2^\pm} = \frac{v_\eta^2 + v_\chi^2}{2v_\eta v_\chi}\left(fv_\rho - 2\lambda_8v_\eta v_\chi\right), \qquad m^2_{H^{\pm\pm}} = \frac{v_\rho^2 + v_\chi^2}{2v_\rho v_\chi}\left(fv_\eta - 2\lambda_9v_\rho v_\chi\right), \\ 
&& m^2_W = \frac{1}{2}\left(\frac{ev_W}{s_W}\right)^2, \qquad m_V^2 = \left(\frac{e}{s_W}\right)^2\frac{v_\eta^2 + v_\chi^2}{2}, \qquad m^2_U = \left(\frac{e}{s_W}\right)^2\frac{v_\rho^2 + v_\chi^2}{2}, \\ 
&& m_Z^2 \approx \left(\frac{ev_\eta}{s_W}\right)^2\frac{1}{2\left(1 - s_W\right)}, \quad m_{Z^\prime}^2 \approx \left(\frac{ev_\chi}{s_W}\right)^2\frac{2\left(1 - s_W^2\right)}{3\left(1 - 4s_W^2\right)}.
\end{eqnarray}\end{subequations}
In the calculations of the Ref. \cite{TO96} the following conditions in imposed:
\begin{equation}
\lambda_4 \approx 2\frac{\lambda_2v_\rho^2 -
\lambda_1v_\eta^2}{v_\eta^2 - v_\rho^2}, \quad \lambda_5v_\eta^2 +
2\lambda_6v_\rho^2 \approx -\frac{v_\eta v_\rho}{2}.
\end{equation}
From the Lagrangean (\ref{llep}) we can see that $m_e, m_\mu, m_\tau \propto
v_\rho$ and $m_E, m_M, m_T \propto v_\chi$. Concerning the ordinary
quarks the masses can be obtained from the Lagrangean (\ref{lqua}) taking
into account $m_u \ll m_c \ll m_t$ and $m_d \ll m_s \ll m_b$.
Therefore, we have 
\begin{subequations}
\begin{eqnarray}
&& m_u = \frac{{\cal G}_1^uv_\eta v_\rho}{{\cal G}_2^uv_\rho + {\cal
G}_3^uv_\eta}, \qquad m_c = \frac{{\cal G}_1^cv_\rho + {\cal
G}_2^cv_\eta}{{\cal G}_3^cv_\eta + {\cal G}_4^cv_\rho}v_\rho, \qquad
m_t = {\cal G}_1^tv_\eta + {\cal G}_2^tv_\rho, \label{massf1}\\ 
&& m_d = \frac{{\cal G}_1^dv_\rho v_\eta}{{\cal G}_2^dv_\eta + {\cal
G}_3^dv_\rho}, \qquad m_s = \frac{{\cal G}_1^sv_\eta + {\cal
G}_2^sv_\rho}{{\cal G}_3^sv_\rho + {\cal G}_4^sv_\eta}v_\eta, \qquad
m_b = {\cal G}_1^bv_\rho + {\cal G}_2^bv_\eta, \label{massf2}
\end{eqnarray}\label{massq}\end{subequations} 
and from the Lagrangean (\ref{yukj}) the heavy quark masses are $m_{J_1}, m_{J_2}, m_{J_3} \propto v_\chi$. In Eqs.
(\ref{massq}) the parameters ${\cal
G}^q_\beta$, $q = u, c, t, d, s, b$ and $\beta = 1, 2, 3,
4$ are functions of the Yukawa couplings.
\newpage


\begin{figure}[h]
\includegraphics{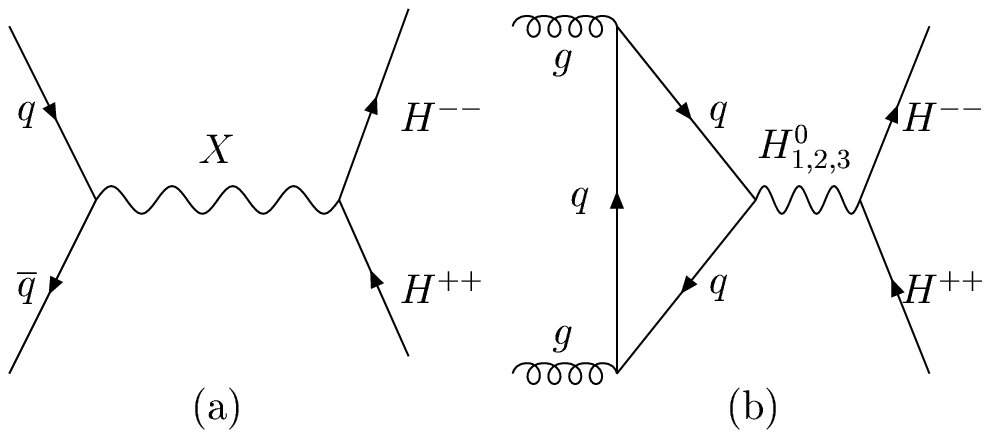}
\end{figure}

\begin{figure}[h]
\caption{\label{fig1} Feynman diagrams for production of doubly charged Higgs bosons {\it via} (a) Drell-Yan process, where $X = \gamma, Z, Z^\prime, H^0_1, H^0_2$ and (b) gluon-gluon fusion.}
\end{figure}

\begin{figure}[h]
\includegraphics{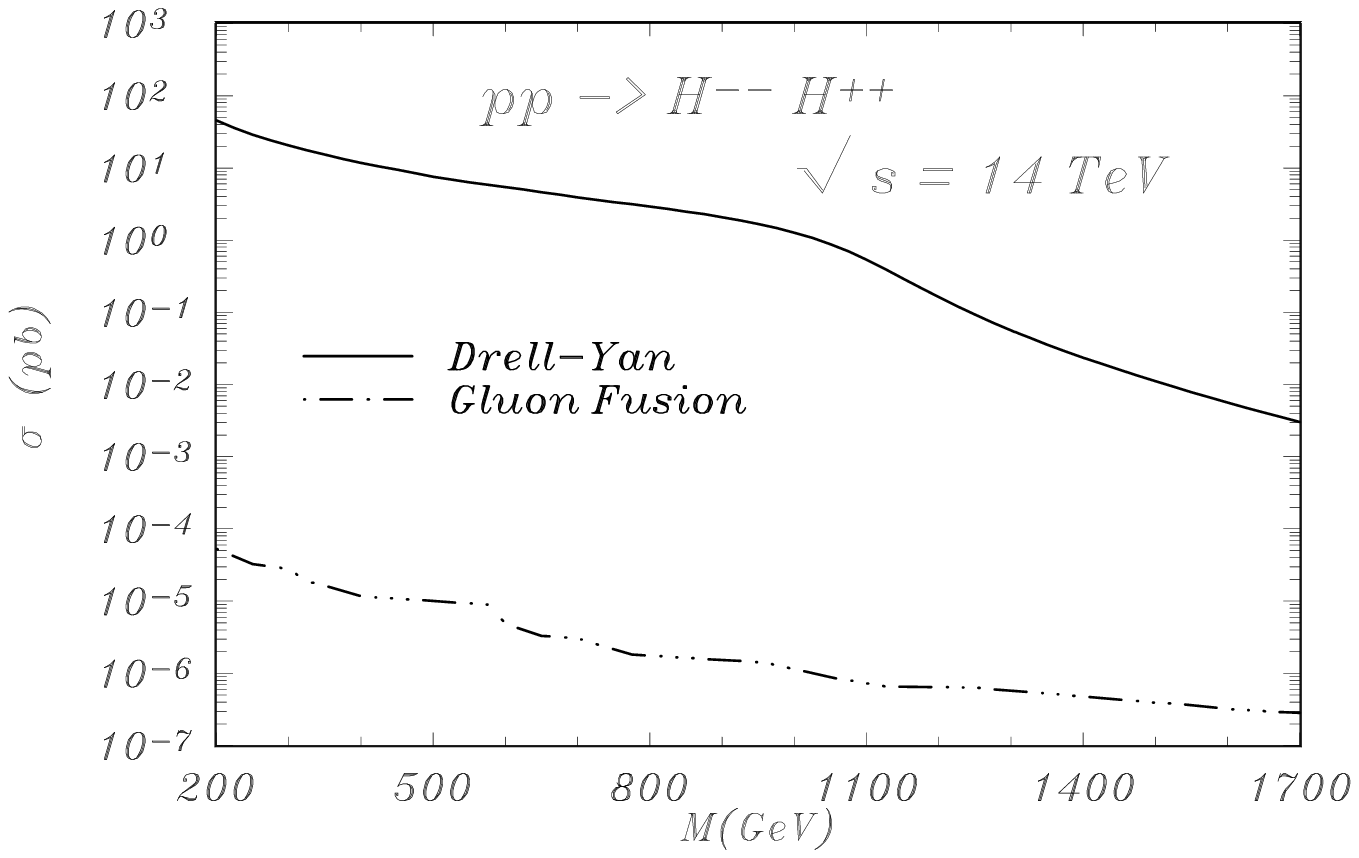}
\caption{\label{fig2} Total cross section for the process $pp \to H^{--}H^{++}$ as a function of $m_{H^{\pm\pm}}$ for $f = 0$ GeV at $\sqrt{s} = 14000$ GeV for Drell-Yan (solid line) and Gluon-Gluon fusion (dot-dash line).}
\end{figure}

\begin{figure}[h]
\includegraphics{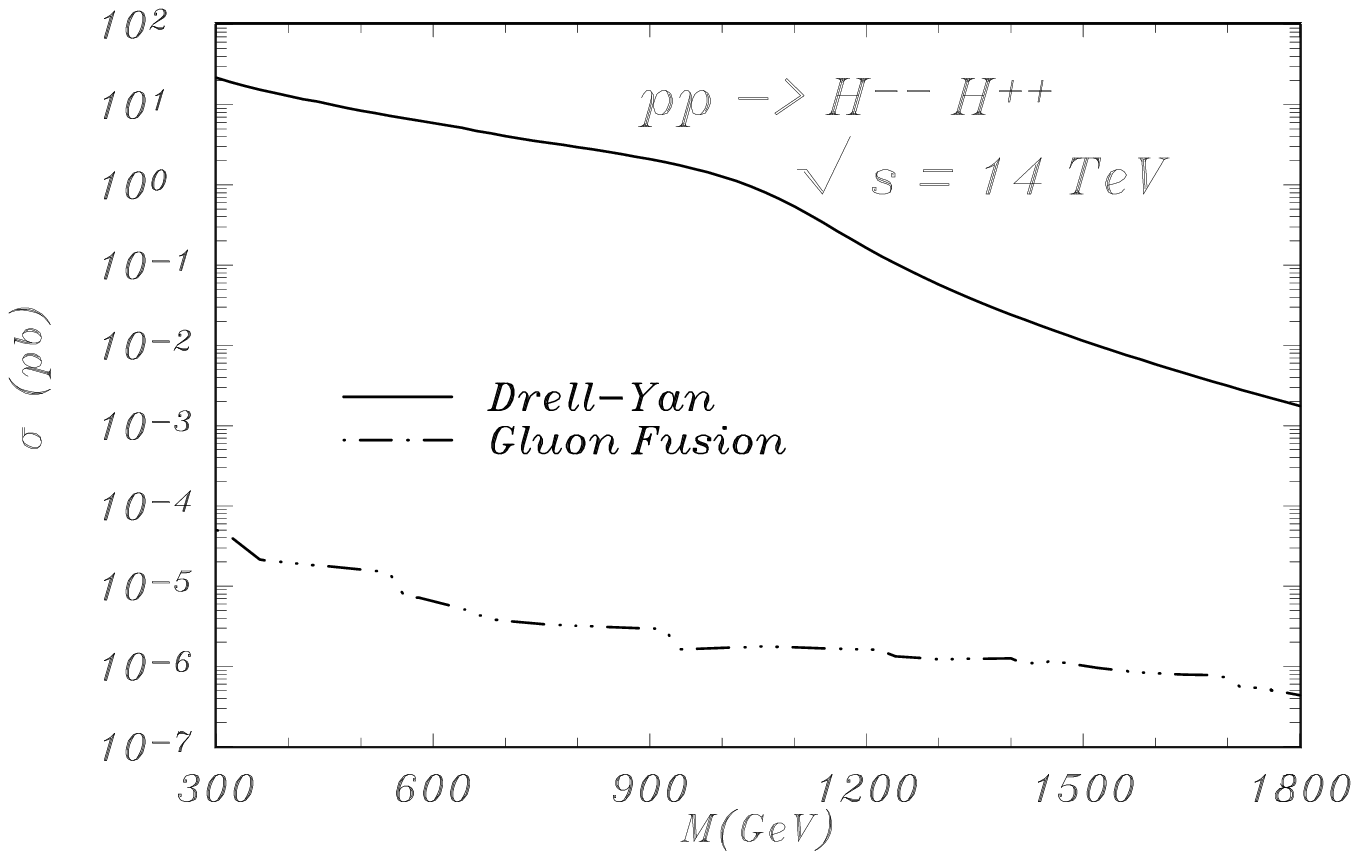}
\caption{\label{fig3}Total cross section for the process $pp \to H^{--}H^{++}$ as a function of $m_{H^{\pm\pm}}$ for $f = -130$ GeV at $\sqrt{s} = 14000$ GeV for Drell-Yan (solid line) and Gluon-Gluon fusion (dot-dash line).}
\end{figure}

\end{document}